\documentclass[a4paper,11pt,nohyper]{JHEP}
\usepackage{amsmath,amssymb}
\usepackage{graphicx}
\usepackage{bm}
\usepackage{booktabs}
\makeatletter
\let\JHEPhash\hash
\newcommand{\JHEPexthref}[2]{%
  \leavevmode
  \pdfstartlink user{%
    /Subtype/Link/Border[0 0 0]/A<</Type/Action/S/URI/URI(#1)>>%
  }%
  \pdfliteral{0 0 0.8 rg 0 0 0.8 RG}%
  #2%
  \pdfliteral{0 0 0 rg 0 0 0 RG}%
  \pdfendlink}
\newcommand{\JHEPgotohref}[2]{% #1 = destination name (tokens after \hash); #2 = link text
  \leavevmode
  \pdfstartlink attr{/Subtype/Link/Border[0 0 0]} goto name{#1}%
  \pdfliteral{0 0 0.8 rg 0 0 0.8 RG}%
  #2%
  \pdfliteral{0 0 0 rg 0 0 0 RG}%
  \pdfendlink}
\def\JHEPnil{}
\renewcommand{\href}[2]{\JHEPhreftest#1\JHEPnil{#1}{#2}}
\def\JHEPhreftest#1#2\JHEPnil#3#4{%
  \ifx#1\JHEPhash
    \JHEPgotohref{#2}{#4}%
  \else
    \JHEPexthref{#3}{#4}%
  \fi}
\renewcommand{\name}[1]{\@ifundefined{pdfoutput}{}{\pdfdest name{#1} xyz}}
\providecommand{\url}[1]{\href{#1}{\texttt{#1}}}
\makeatother

\newcommand{\half}{{\textstyle\tfrac{1}{2}}}
\newcommand{\quarter}{{\textstyle\tfrac{1}{4}}}
\newcommand{\be}{\begin{equation}}
\newcommand{\ee}{\end{equation}}
\newcommand{\ba}{\begin{array}}
\newcommand{\ea}{\end{array}}
\newcommand{\rd}{\mathrm{d}}
\newcommand{\mm}{\mathfrak{m}}
\newcommand{\qq}{\mathfrak{q}}
\newcommand{\mmj}{\bar{\mathfrak{m}}_{j}}
\newcommand{\rBL}{r_{\rm BL}}
\newcommand{\betappn}{\beta_{{\rm PPN}}}
\newcommand{\gammappn}{\gamma_{{\rm PPN}}}
\newcommand{\cB}{\mathcal{B}}
\newcommand{\cF}{\mathcal{F}}
\newcommand{\cG}{\mathcal{G}}
\newcommand{\cH}{\mathcal{H}}

\newcommand{\SO}{\mathbf{SO}}
\newcommand{\SL}{\mathbf{SL}}
\newcommand{\ODD}{\mathbf{O}(D,D)}

\title{Exact Four-Parameter Rotating NS--NS Vacuum in Double Field Theory}

\author{Hun Jang,$^{a,b}$ Minkyoo Kim,$^{c,d}$ Jeong-Hyuck Park$^{a}$\\

$^{a}$Department of Physics, Sogang University, Seoul 04107, Korea\\
$^{b}$Yukawa Institute for Theoretical Physics, Kyoto University, Kyoto 606-8502, Japan\\
$^{c}$Department of Physics and Astronomy, Seoul National University, Seoul 08826, Korea\\
$^{d}$School of Science, Huzhou Normal University, Huzhou 313000, China\\

E-mail: \email{hun.jang@nyu.edu}, \email{minkyookim@snu.ac.kr}, \email{park@sogang.ac.kr}}

\abstract{%
We construct an exact rotating vacuum of the NS--NS sector, equivalently a Double Field Theory vacuum.
The construction applies compact $\mathbf{SO}(2)$ S-duality to a rotating Einstein--scalar seed.
The solution has four independent parameters $\{\mathfrak{m},j,\mathfrak{q},\zeta\}$.
To our knowledge, this is the first explicit rotating solution of the pure NS--NS vacuum equations with all three NS--NS fields $\{g,B,\phi\}$ determined analytically, independent dilaton and $H$-flux charges, and no Maxwell sector.
The static limit is obtained by taking the rotation parameter $j\to 0$.
In this limit the geometry is not the spherical Burgess--Myers--Quevedo solution.
Instead, it is an axial Zipoy--Voorhees branch carrying $H$-flux, so an oblate deformation remains after rotation is switched off.
This geometric memory is absent in pure general relativity and in Einstein--Maxwell--dilaton--axion.
The two static branches nevertheless share the same $\ell=0$ parametrized post-Newtonian data $\{MG,\beta_{\mathrm{PPN}},\gamma_{\mathrm{PPN}},h\}$.
They give two inequivalent NS--NS geometries at identical monopole charges, with the degeneracy lifted at $\ell=2$.
At the Kerr horizon locus the outer shell is generically singular in curvature.
Above the threshold $|\mathfrak{q}|>\sqrt{\mathfrak{m}^{2}-j^{2}}$, polar geodesics are repelled outward and the rotation axis becomes regular in curvature at the shell.
On that axis the inverse metric $g^{\mu\nu}$ stays finite while the lower-index Riemannian metric components diverge, whereas off the axis the inverse metric itself diverges.
This axis-local degeneracy may offer a setting for non-Riemannian geometry in Double Field Theory, where $g_{\mu\nu}$ is not fundamental and the $\mathbf{O}(D,D)$ variables $\{d,\mathcal{H}_{AB}\}$ remain well defined.}


\begin{document}

\setcounter{tocdepth}{2}
\tableofcontents

\vspace{-5pt}

%======================== BODY ========================

\section{Introduction}
The Kerr solution is the canonical stationary rotating vacuum of four-dimensional General Relativity (GR), confirmed by gravitational wave detections~\cite{LIGOScientific:2016aoc} and horizon-scale imaging~\cite{EventHorizonTelescope:2019dse,EventHorizonTelescope:2022wkp}, and uniquely fixed in the electrovacuum sector by the three asymptotic charges $\{M,J,Q\}$~\cite{Israel:1967wq,Carter:1971zc,Hawking:1972qk,Robinson:1975bv}.  Closed string theory, however, always contains the NS--NS triplet $\{g_{\mu\nu},B_{\mu\nu},\phi\}$ in its massless sector~\cite{Green:1987sp,Callan:1985ia}; throughout, $g_{\mu\nu}$ denotes the string frame metric, and the four-dimensional NS--NS vacuum equations
\be
\ba{l}
R_{\mu\nu}+2\nabla_{\mu}\partial_{\nu}\phi-\quarter H_{\mu\rho\sigma}H_{\nu}{}^{\rho\sigma}=0\,,\\[2pt]
\half e^{2\phi}\nabla^{\rho}\!\left(e^{-2\phi}H_{\rho\mu\nu}\right)=0\,,\\[2pt]
R+4\Box\phi-4\partial_{\mu}\phi\,\partial^{\mu}\phi-{\textstyle\frac{1}{12}}H_{\lambda\mu\nu}H^{\lambda\mu\nu}=0
\ea
\label{EDFE}
\ee
are taken at $\alpha'$-leading order.  In Double Field Theory (DFT)~\cite{Hull:2009mi,Hohm:2010pp} this triplet is itself the gravitational content, packaged into the $\ODD$-singlet dilaton $d$ and the generalized metric $\cH_{AB}$.  In the absence of external matter, the $\ODD$-covariant DFT Einstein equation reduces to the vacuum equation~\cite{Angus:2018mep}
\be
G_{AB}=0\,.
\label{DFTEvac}
\ee
Under a Riemannian parametrization, this becomes~(\ref{EDFE}) (see~\cite{Park:2025ugx} for a recent review of the DFT Einstein equation).

Once an additional $\mathbf{U}(1)$ Maxwell sector is coupled, the Hassan--Sen $\mathbf{O}(d,d)$ twist on Kerr returns Kerr--Sen~\cite{Sen:1992ua}, and the Einstein--Maxwell--dilaton--axion (EMDA) family of Garc\'{\i}a--Gal'tsov--Kechkin~\cite{Garcia:1995qz} and that of Gal'tsov--Karsanov~\cite{Galtsov:2025nia} extend Kerr--Sen.  In all three cases the axidilaton charge is algebraically locked to the Maxwell charge; these are solutions of the enlarged NS--NS\,+\,Maxwell system, not of~(\ref{EDFE}) alone.  To our knowledge, no explicit rotating solution of the pure NS--NS vacuum equations~(\ref{EDFE}) with independent dilaton and $H$-flux charges and no Maxwell sector was previously available.  We construct such a solution: an analytic NS--NS vacuum with four parameters, namely bare mass $\mm$, rotation $j$, $\SO(2)$-invariant scalar charge $\qq$, and compact $\SO(2)$ S-duality angle $\zeta$ rotating the dilaton into the axion.  The construction applies the compact $\SO(2)\!\subset\!\SL(2,\mathbb{R})$ S-duality that preserves asymptotic flatness to the rotating Einstein--scalar Bogush--Gal'tsov seed~\cite{Bogush:2020lkp} (the rotating extension of the Fisher--Janis--Newman--Winicour (FJNW)--Wyman solution~\cite{Fisher:1948,Janis:1968,Wyman:1981}), then performs the Weyl rescaling to string frame and Hodge dualizes the axion to a Kalb--Ramond two-form.

We emphasize three structural features of the solution.  (a) \emph{The static limit is axial, not spherical Burgess--Myers--Quevedo (BMQ)}: the ${j\to 0}$ limit lands on an axial Zipoy--Voorhees (ZV) geometry carrying $H$-flux~\cite{Zipoy:1966,Voorhees:1970}, \emph{not} the spherical BMQ solution~\cite{Burgess:1994kq}; an oblate Weyl class deformation persists in the limit, a geometric memory absent in pure GR and EMDA.  (b) \emph{Monopole degeneracy despite axial geometry}: this branch and spherical BMQ share identical monopole and parametrized post-Newtonian (PPN) data $\{MG,\betappn,\gammappn,h\}$, the degeneracy lifting only at the next multipole; although $\phi$ and $H$-flux are $\SO(3)$ invariant, the metric $g_{\mu\nu}$ and derived DFT variables $\{d,\cH_{AB}\}$ are only axial $\mathbf{U}(1)$.  (c) \emph{Polar repulsion}: above the threshold $|\qq|>\sqrt{\mm^{2}-j^{2}}$ ($\sin 2\zeta\neq 0$), polar geodesics are repelled outward, reflecting the dilaton's negative kinetic term in string frame.  We also note a closed-form $\SO(2)_{\zeta}$-invariant first integral and a stringy Lense--Thirring correction at $1/r^{2}$ set by the dilaton charge alone.\vspace{5pt}

\section{The solution in the quasi-isotropic frame}
The solution is labeled by
\be
\{\,\mm,\;j,\;\qq,\;\zeta\,\}\,,
\label{parameters}
\ee
with admissible ranges $\mm>0$, $|j|<\mm$, $\qq\in\mathbb{R}$, $\zeta\in[0,\pi/2)$; the displayed formulas are subextremal, $|j|=\mm$ being the smooth $\mmj\to 0$ limit.  Throughout this paper, $r$ denotes the quasi-isotropic radial coordinate; the Boyer--Lindquist coordinate $\rBL$ enters only as an auxiliary function of $r$.  In coordinates $(t,r,\vartheta,\varphi)$, we define
\be
\ba{ll}
\mmj\equiv\sqrt{\mm^{2}-j^{2}}\,,\quad&\quad
\rBL(r)=r+\mm+\dfrac{\mmj^{2}}{4r}\,,\\
\cG(r)\equiv\dfrac{2r-\mmj}{2r+\mmj}\,,\quad&\quad
\sqrt{\Delta(r)}=r-\dfrac{\mmj^{2}}{4r}\,,\\
\Sigma_{\pm}(r,\vartheta)\equiv\rBL^{2}(r)\pm j^{2}\cos^{2}\!\vartheta\,,\quad&\quad
\Lambda(r,\vartheta)\equiv\Delta(r)-j^{2}\sin^{2}\!\vartheta\,.
\ea
\label{aux}
\ee
The outer shell sits at $r_{+}\!=\!\mmj/2$, where $\cG\to 0$ and $\Delta\to 0$.  The dilaton is purely radial,
\be
\boxed{e^{2\phi(r)}=\cG(r)^{2\qq/\mmj}\cos^{2}\!\zeta+\cG(r)^{-2\qq/\mmj}\sin^{2}\!\zeta\,,}
\label{dilaton}
\ee
the Kalb--Ramond field has a single nonzero component, $B_{t\varphi}=h\cos\vartheta$ with $h\equiv-2\qq\sin 2\zeta$, so the three-form $H$-flux is purely electric,
\be
\boxed{H_{(3)}=h\sin\vartheta\,\rd t\wedge\rd\vartheta\wedge\rd\varphi\,,}
\label{Hflux}
\ee
and the string frame line element takes the Kerr form dressed by the dilaton,

\be
\boxed{
\rd s^{2}=e^{2\phi(r)}\!\left[-\dfrac{\Lambda(r,\vartheta)}{\Sigma_{+}(r,\vartheta)}\big(\rd t-\omega(r,\vartheta)\,\rd\varphi\big)^{2}+\cB(r,\vartheta)\big(\rd r^{2}+r^{2}\rd\vartheta^{2}\big)+\dfrac{\Sigma_{+}(r,\vartheta)\,\Delta(r)\sin^{2}\!\vartheta}{\Lambda(r,\vartheta)}\rd\varphi^{2}\right],}
\label{metric}
\ee

with conformal warp factor and frame-dragging function
\be
\ba{ll}
\cB(r,\vartheta)=\dfrac{\Sigma_{+}(r,\vartheta)}{r^{2}}\!\left(1+\dfrac{\mmj^{2}\sin^{2}\!\vartheta}{\Delta(r)}\right)^{\!-\qq^{2}/\mmj^{2}}\,,\qquad&\quad
\omega(r,\vartheta)=-\dfrac{2\mm j\,\rBL(r)\sin^{2}\!\vartheta}{\Lambda(r,\vartheta)}\,.
\ea
\label{Bomega}
\ee
The dilaton~(\ref{dilaton}) is the exact rotating generalization of the static BMQ dilaton~\cite{Burgess:1994kq} in the isotropic coordinate of~\cite{Choi:2022fqj}: rotation enters only through the replacement $\mm\to\mmj$.  The factor $\cB(r,\vartheta)$ in the $(r,\vartheta)$-block carries the scalar hair $\qq$; its angular factor $(1+\mmj^{2}\sin^{2}\!\vartheta/\Delta)^{-\qq^{2}/\mmj^{2}}$ reduces to unity in the Kerr limit $\qq\to 0$, leaving $\cB\to\Sigma_{+}/r^{2}$.\vspace{5pt}
\begin{figure}[!htbp]
\centering
\includegraphics[width=0.62\linewidth]{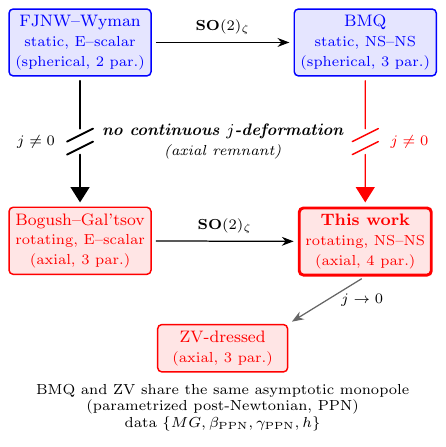}
\caption{\label{fig:sduality}Solution-generating square.  The compact S-duality $\SO(2)_{\zeta}\!\subset\!\SL(2,\mathbb{R})$ maps Einstein--scalar seeds~\cite{Fisher:1948,Janis:1968,Wyman:1981,Bogush:2020lkp} to NS--NS descendants~\cite{Burgess:1994kq}.  For $\qq\neq 0$, the rotating branch has a ZV branch carrying $H$-flux as its static limit.  It shares the BMQ monopoles but not the spherical BMQ geometry; the broken vertical arrows mark this missing return to the spherical row.}
\end{figure}

\section{Construction and conservation law}
At tree level and to leading order in $\alpha'$, the four-dimensional NS--NS sector $\{g_{\mu\nu},B_{\mu\nu},\phi\}$ obeys~(\ref{EDFE}).  In double field theory these equations reassemble into the $\ODD$ covariant vacuum generalized Einstein equation~(\ref{DFTEvac})~\cite{Angus:2018mep,Park:2025ugx} on the pair $\{d,\cH_{AB}\}$ of $\ODD$ singlet dilaton $d=\phi-\half\ln\sqrt{-g}$ and generalized metric
\be
\cH_{AB}(g,B)=\begin{pmatrix}g^{\mu\nu} & -g^{\mu\rho}B_{\rho\nu}\\ B_{\mu\rho}g^{\rho\nu} & g_{\mu\nu}-B_{\mu\rho}g^{\rho\sigma}B_{\sigma\nu}\end{pmatrix}\,,
\label{eq:cH}
\ee
which places $\{g,B\}$ on equal footing.  Equation~(\ref{eq:cH}) is the Riemannian parametrization of the DFT generalized metric.  More generally, the fundamental fields are $d$ and $\cH_{AB}$ themselves, and $\cH_{AB}$ also admits non-Riemannian parametrizations for which no Riemannian metric $g_{\mu\nu}$ can be defined locally~\cite{Morand:2017fnv}.

We start from the Bogush--Gal'tsov rotating Einstein--scalar solution~\cite{Bogush:2020lkp}, generated by a Cl\'ement transformation of FJNW--Wyman~\cite{Fisher:1948,Janis:1968,Wyman:1981}.  This seed was recently rederived using the Eri\c{s}--G\"{u}rses integrability framework~\cite{Barrientos:2025rot}.  The Einstein-frame seed line element is
\be
\rd s^{2}_{E}=-\dfrac{\Lambda}{\Sigma_{+}}(\rd t-\omega\rd\varphi)^{2}+\cB(r,\vartheta)(\rd r^{2}+r^{2}\rd\vartheta^{2})+\dfrac{\Sigma_{+}\,\Delta\sin^{2}\!\vartheta}{\Lambda}\rd\varphi^{2}\,,
\label{eq:BG_E_metric}
\ee
with $\omega,\cB$ as in~(\ref{Bomega}), seed dilaton $\phi_{\rm seed}=(\qq/\mmj)\ln\cG$, and vanishing axion.  This pair solves $R^{E}_{\mu\nu}=2\partial_{\mu}\phi_{\rm seed}\partial_{\nu}\phi_{\rm seed}$ and $\nabla_{E}^{2}\phi_{\rm seed}=0$, with sphericity broken to axial $\mathbf{U}(1)$ at $j=0$ via $\cB(r,\vartheta)$.  With the auxiliary functions of~(\ref{aux}), the Boyer--Lindquist combination collapses to a perfect square,
\be
\dfrac{\rBL-\mm-\mmj}{\rBL-\mm+\mmj}=\left(\dfrac{2r-\mmj}{2r+\mmj}\right)^{\!2}=\cG(r)^{2}\,,
\label{eq:perfsq}
\ee
the main simplification afforded by the quasi-isotropic chart.

The complex axidilaton is $\lambda=\chi+ie^{-2\phi}$ ($\mathrm{Im}\,\lambda>0$), parametrizing $\SL(2,\mathbb{R})/\SO(2)$.  The standard compact subgroup $\SO(2)_{\zeta}\!\subset\!\SL(2,\mathbb{R})$ preserving asymptotic flatness acts by $\lambda\to\lambda'=(\cos\zeta\,\lambda+\sin\zeta)/(-\sin\zeta\,\lambda+\cos\zeta)$, $\zeta\in[0,\pi/2)$.  Applying it to $\lambda_{\rm seed}=i\cG^{-2\qq/\mmj}$, rationalizing, and reading $\lambda'=\chi'+ie^{-2\phi'}$ gives
\be
e^{2\phi'(r)}=\cG^{2\qq/\mmj}\cos^{2}\!\zeta+\cG^{-2\qq/\mmj}\sin^{2}\!\zeta\,,\quad
\chi'(r)=\dfrac{\half\sin 2\zeta\,(\cG^{2\qq/\mmj}-\cG^{-2\qq/\mmj})}{e^{2\phi'(r)}}\,,
\label{eq:rotated}
\ee
i.e.\ the boxed dilaton~(\ref{dilaton}).  The Einstein-frame metric is an $\SL(2,\mathbb{R})$ singlet and is unchanged.

The string frame metric is $g^{\rm str}_{\mu\nu}=e^{2\phi'}g^{E}_{\mu\nu}$, yielding the line element~(\ref{metric}).  In four dimensions the closed-string axion is dual to the NS--NS three-form, $H_{(3)}=e^{2\phi'}\,{\star}_{\rm str}\,\rd\chi'$; with $\chi'$ depending only on $r$ the explicit Hodge dual evaluates to the $H$-flux~(\ref{Hflux}), $H_{(3)}=h\sin\vartheta\,\rd t\wedge\rd\vartheta\wedge\rd\varphi$ with $h=-2\qq\sin 2\zeta$ and local Kalb--Ramond potential $B_{t\varphi}=h\cos\vartheta$.  The same construction applied to the static FJNW--Wyman seed reproduces the spherical BMQ solution~\cite{Burgess:1994kq} along the chain from FJNW--Wyman to BMQ (Fig.~\ref{fig:sduality}).  A previous complexification in the Newman--Janis style of the spherical $D=4$ DFT vacuum~\cite{Angus:2018mep} obtained a rotating metric~\cite{Li:2024ijj} but did not determine the dilaton or $H$-flux explicitly in the rotating sector; the present $\SL(2,\mathbb{R})/\SO(2)$ S-duality construction yields all NS--NS fields in closed form.  As a direct check, Appendix~\ref{app:BL} rewrites the configuration in Boyer--Lindquist coordinates; the corresponding Maple/GRTensorIII and Mathematica verification of the vacuum equations~(\ref{EDFE}) is archived in~\cite{Code:GitHub}.

The compact $\SO(2)_{\zeta}$ transformation used here should not be confused with an $\ODD$ transformation of the NS--NS generalized metric.  It acts on the four-dimensional axidilaton after dualizing the two-form to an axion, whereas $\ODD$ T-duality acts on $\cH_{AB}$ with the DFT dilaton $d$ as a singlet.  The Hodge star used in the axion/$H$-flux dualization depends on a four-dimensional Riemannian parametrization, in particular on the metric and volume form; it is therefore not an $\ODD$-covariant operation on $\cH_{AB}$ and $d$ alone.  Thus our construction is a solution-generating map between Riemannian parametrizations of the DFT fields, rather than a single $\ODD$ rotation.  A formulation in which the $\SL(2)$ and $\ODD$ actions are manifest simultaneously would require an enlarged extended-field-theory framework, such as the $\SL(2)\times\mathbf{O}(6,6+n)$ construction of~\cite{Ciceri:2016xcm}.

The radial dilaton equation integrates once to the closed-form first integral
\be
\boxed{\;(2\Delta\,\partial_{\rBL}\phi)^{2}+h^{2}\,e^{-4\phi}=4\qq^{2}\,.}
\label{eq:firstInt_body}
\ee
This is the $\SO(2)_{\zeta}$-invariant first integral of the axidilaton, conserved on the circle of radius $2|\qq|$.  The $\SO(2)_{\zeta}$ orbit rotates the dilaton kinetic charge into the dual axion charge, equivalently the $H$-flux charge in the NS--NS frame, while the right-hand side remains fixed at the invariant radius $4\qq^{2}$.\vspace{5pt}

\section{Properties of the solution}
\label{sec:properties}

\subsection{Geometric memory: the static limit is axial, not spherical}
\begin{figure}[h]
\centering
\includegraphics[width=0.6\linewidth]{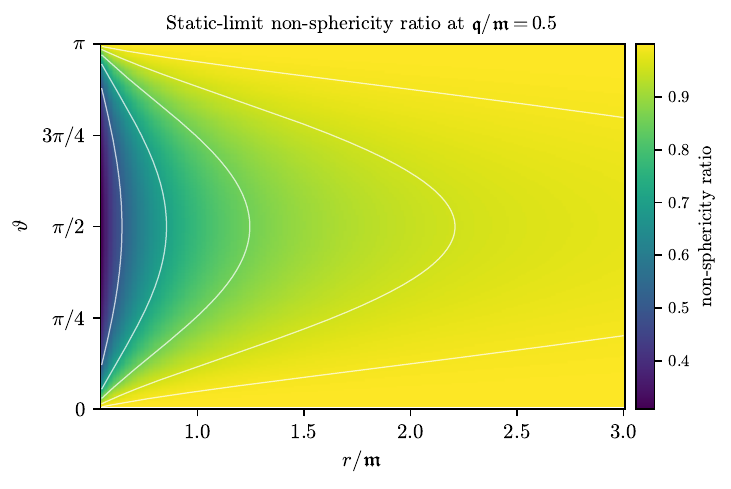}
\caption{\label{fig:s2nonsphere}Plot of the static-limit non-sphericity ratio, Eq.~(\ref{nonspherical}), over $(r/\mm,\vartheta)$ at $\qq/\mm=0.5$.  The ratio departs from unity for any $\qq\neq 0$ except on the rotation axis ($\vartheta=0,\pi$): the static $j\to 0$ limit lies in the Zipoy--Voorhees class, not on the spherical BMQ branch.}
\end{figure}

The $j\to 0$ limit of~(\ref{metric}) is \emph{not} the spherical BMQ solution.  At $j=0$ one has $\mmj\to\mm$, yet $\cB(r,\vartheta)$ retains $\vartheta$-dependence for any $\qq\neq 0$:
\be
\frac{g_{\vartheta\vartheta}}{g_{\varphi\varphi}/\sin^{2}\!\vartheta}\bigg|_{j=0}=\!\left(1+\dfrac{\mm^{2}\sin^{2}\!\vartheta}{\Delta}\right)^{\!-\qq^{2}/\mm^{2}}\!,
\label{nonspherical}
\ee
departing from unity at every generic $\vartheta$ (equality only on the axis $\sin\vartheta=0$; Fig.~\ref{fig:s2nonsphere}).  The surface at fixed $(t,r)$ is therefore not a round sphere.  The static limit lies in the ZV class dressed by closed-string fields~\cite{Zipoy:1966,Voorhees:1970}: the ZV metric augmented by the NS--NS dilaton and by $H$-flux, the latter equivalently described in four dimensions by the dual axion.  This is an oblate Weyl class deformation inherited from the Einstein--scalar seed (Fig.~\ref{fig:sduality}).  The obstruction is absent from both pure GR, where Kerr reduces to Schwarzschild and the $j\to 0$ limit preserves exact sphericity, and EMDA, where the Killing tensor algebra carries sphericity through to Kerr--Sen~\cite{Sen:1992ua} and the Garc\'{\i}a--Gal'tsov--Kechkin and Gal'tsov--Karsanov (GGK/GK) families~\cite{Garcia:1995qz,Galtsov:2025nia}.  It is intrinsic to the NS--NS scalar hair sector.  Rotation thus permanently imprints the spatial geometry: the oblate closed-string-dressed ZV deformation survives the $j\to 0$ limit as a residual feature of the closed-string hair.\vspace{5pt}

Known constructions in the pure NS--NS sector appear to encounter the same axial remnant.  The conformal Kerr ansatz violates the Einstein-frame scalar equation, the Newman--Janis procedure fails on the Fisher seed~\cite{Bogush:2020lkp}, and the Hassan--Sen twist requires a Maxwell sector absent from~(\ref{EDFE}).  Thus the present ZV branch carrying $H$-flux emerges naturally within the currently known NS--NS toolkit.

\subsection{Monopole degeneracy and higher multipole lifting}
Expanding the string frame metric and dilaton at large $r$ in the PPN form
\be
\ba{rll}
g_{tt}&\simeq&-1+\dfrac{2MG}{r}-\dfrac{2\betappn(MG)^{2}}{r^{2}}\,,\\
g_{rr}&\simeq& 1+\dfrac{2\gammappn MG}{r}\,,\\
e^{2\phi}&\simeq& 1+\dfrac{(\gammappn-1)MG}{r}\,,
\ea
\label{PPN}
\ee
the asymptotic charges read
\be
\ba{llll}
MG=\mm+\qq\cos 2\zeta\,,\quad&\quad
\betappn=1+\dfrac{\qq^{2}\sin^{2}\!2\zeta}{(MG)^{2}}\,,\quad&\quad
\gammappn=\dfrac{\mm-\qq\cos 2\zeta}{\mm+\qq\cos 2\zeta}\,,\quad&\quad J=\mm j\,.
\ea
\label{MGbetagamma}
\ee
The former three quantities are independent of $j$ at leading order and identical to those of the spherical BMQ branch~\cite{Choi:2022fqj}.  Thus the same monopole/PPN data $\{MG,\betappn,\gammappn,h\}$ are realized by two inequivalent static NS--NS vacuum geometries: an $\ell=0$ asymptotic degeneracy that lifts only at $\ell=2$ through the curvature invariants and the next multipole order.  Writing $b$ for the scalar-charge parameter of the static BMQ family, the BMQ reality condition $b^{2}\geq h^{2}$ becomes $4\qq^{2}\geq h^{2}$ in the present dictionary and is automatic for $h=-2\qq\sin2\zeta$; the $h=0$ slice is the FJNW/dilaton-only subbranch, not a restriction on BMQ.  The scalar $\phi$ and the gauge-invariant three-form $H_{(3)}$ are $\SO(3)$ invariant, while $B_{(2)}$ realizes the symmetry only modulo gauge transformations, as expected for a Dirac-monopole-type potential; by contrast, the string frame metric is axial.  In DFT, however, the natural variables are the $\ODD$ singlet dilaton $d=\phi-\half\ln\sqrt{-g}$ and the generalized metric $\cH_{AB}(g,B)$; the singlet
\be
e^{-2d}=e^{2\phi}\,\Sigma_{+}\,\sin\vartheta\,\dfrac{\sqrt{\Delta}}{r}\!\left(1+\dfrac{\mmj^{2}\sin^{2}\!\vartheta}{\Delta}\right)^{\!-\qq^{2}/\mmj^{2}}\,,
\label{DFTdilaton}
\ee
inherits the $\vartheta$-dependence of $\cB$, and $\cH_{AB}$ inherits the axial structure of $g_{\mu\nu}$: both DFT fields break $\SO(3)$ down to axial $\mathbf{U}(1)$.  Expanding the non-sphericity ratio~(\ref{nonspherical}) for large $\rBL$ gives $1-\qq^{2}\sin^{2}\!\vartheta/\rBL^{2}+\mathcal{O}(\rBL^{-3})$, so the ZV branch develops a nonzero $1/\rBL^{2}$ mass quadrupole already at $\mathcal{O}(\qq^{2}/\mm^{2})$ while the spherical BMQ branch contributes none at any $\ell\geq 1$; a coordinate-invariant probe is the Ricci scalar in the static limit~(\ref{eq:Ricci}), dependent on $\vartheta$ on the ZV branch and independent of $\vartheta$ on the BMQ branch.  Such failures of asymptotic uniqueness have analogues in the Zipoy--Voorhees~\cite{Zipoy:1966,Voorhees:1970} and Manko--Novikov~\cite{Manko:1992} geometries, which share leading asymptotic charges while differing at higher multipoles.  At the level of the field equations, with $(\mm,\qq,\zeta)$ matched, both branches yield the same $\{MG,\betappn,\gammappn,h\}$; they differ in the bulk $\cB(r,\vartheta)$, with the difference appearing only at $\ell\geq 2$.\vspace{5pt}

\subsection{Stringy rotational effects}
The stationary rotation of the closed-string hair leaves two distinctive signatures: a frame-dragging correction at post-Newtonian order, and a strong-field polar repulsion near the outer shell.

\subsubsection{Lense--Thirring correction}
Beyond the leading PPN form~(\ref{PPN}), the frame-dragging coefficient $g_{t\varphi}$ acquires a non-Kerr $1/r^{2}$ correction,
\be
g_{t\varphi}=-\dfrac{2\mm j\sin^{2}\!\vartheta}{r}+\dfrac{2\mm j(\mm+2\qq\cos 2\zeta)\sin^{2}\!\vartheta}{r^{2}}+\mathcal{O}(r^{-3})\,,
\label{eq:LT_body}
\ee
displacing the Kerr Lense--Thirring coefficient $2\mm^{2}j$ by $+4\mm j\qq\cos 2\zeta$.  The correction is controlled by the dilaton charge $\qq\cos 2\zeta$ alone (the same combination that enters $\gammappn$), providing an independent dynamical probe through frame dragging; purely axionic hair $\zeta=\pi/4$ leaves the $1/r^{2}$ coefficient at its Kerr value, just as it leaves $\gammappn=1$.\footnote{For the Sun, the Cassini bound on $|\gammappn-1|$~\cite{Bertotti:2003rm} together with the PPN bound on $|\betappn-1|$~\cite{Will:2014kxa} forces $|\qq|/MG\lesssim 10^{-2}$, placing the polar repulsion threshold $|\qq|>\mmj$ beyond solar reach; compact object systems in which $\qq$ is governed by strong-field NS--NS hair may offer more accessible signals.}

Resolving the next $\vartheta$-dependent order, $g_{t\varphi}$ admits the quasi-isotropic asymptotic expansion
\be
\boxed{\,g_{t\varphi}(r,\vartheta)=-\dfrac{2\mm j\sin^{2}\!\vartheta}{r}+\dfrac{2\mm j(\mm+2\qq\cos 2\zeta)\sin^{2}\!\vartheta}{r^{2}}-\dfrac{2\mm j\,\mathcal{K}(\vartheta)\sin^{2}\!\vartheta}{r^{3}}+\mathcal{O}(r^{-4})\,,\,}
\label{eq:LT_expansion}
\ee
where the cubic term is written with the explicit minus sign shown above, and the angular factor is
\be
\mathcal{K}(\vartheta)\equiv 2\qq^{2}+2\mm\qq\cos 2\zeta+\tfrac{3}{4}\mm^{2}+\tfrac{1}{4}j^{2}-j^{2}\cos^{2}\!\vartheta\,.
\label{eq:Kfunc}
\ee
Beyond the $1/r^{2}$ Kerr plus string correction in~(\ref{eq:LT_body}), the cubic angular factor $\mathcal{K}(\vartheta)$ carries the pure Kerr contributions $\tfrac{3}{4}\mm^{2}+\tfrac{1}{4}j^{2}-j^{2}\cos^{2}\!\vartheta$ together with terms induced by hair, $2\qq^{2}+2\mm\qq\cos 2\zeta$, that shift the effective quadrupole moment away from its pure Kerr value.\vspace{5pt}

\begin{figure}[!t]
\centering
\newcommand{\leftShift}{0pt}%
\newcommand{\rightShift}{5em}%
\newcommand{\leftWidth}{0.54\linewidth}%
\newcommand{\rightWidth}{0.42\linewidth}%
\newcommand{\figCaptionGap}{-40pt}%
\raisebox{\leftShift}{\includegraphics[width=\leftWidth]{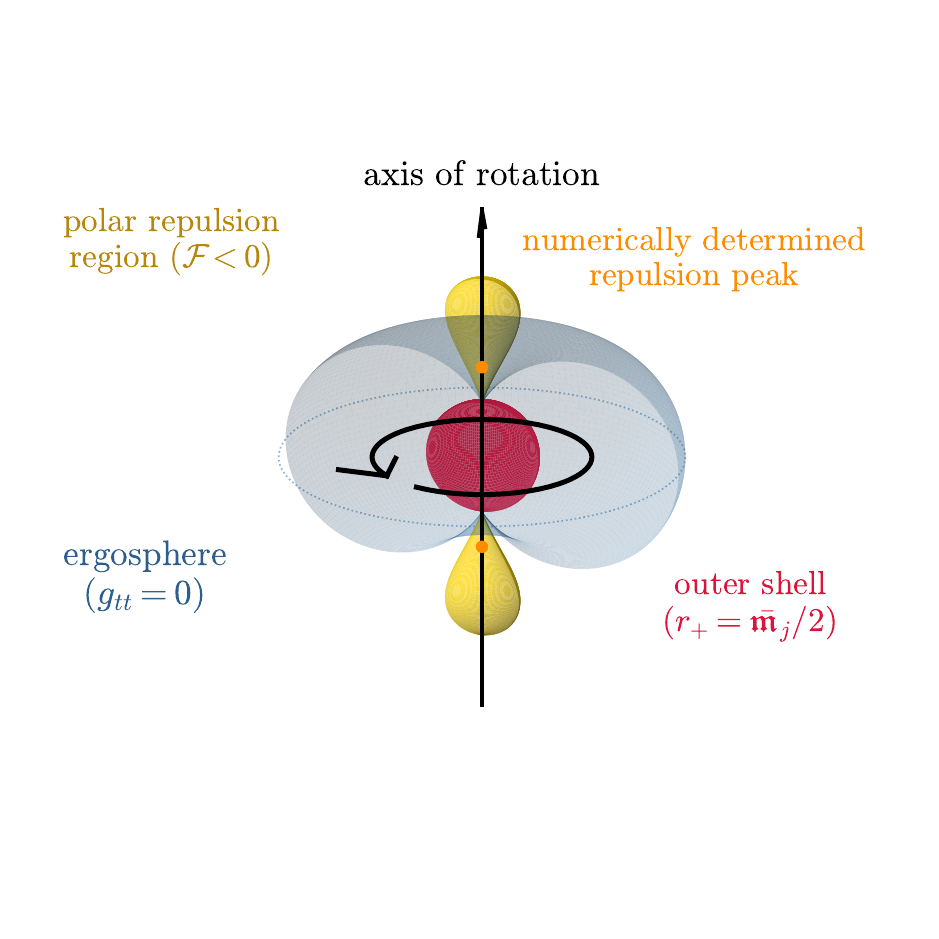}}%
\hfill%
\raisebox{\rightShift}{\includegraphics[width=\rightWidth]{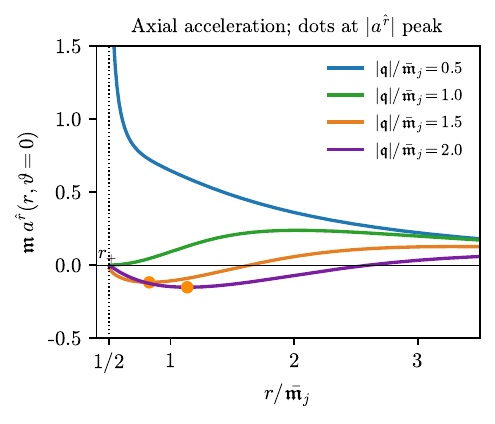}}%
\vspace{\figCaptionGap}
\caption{\label{fig:polar}Polar repulsion at $\mm=1$, $j=0.85$, $\zeta=0.5$.  \textit{Left}: 3D quasi-isotropic embedding at $|\qq|/\mmj=1.5$, with ergosurface (steelblue), outer shell $r_{+}=\mmj/2$ (crimson), and $\cF<0$ regions (gold) wrapping the axis.  \textit{Right}: axial $\mm\,a^{\hat r}(r,0)$ vs $r/\mmj$ for $|\qq|/\mmj\in\{0.5,1.0,1.5,2.0\}$; shell at $r/\mmj=1/2$ (dotted).  Supercritical curves ($|\qq|>\mmj$) approach $0^{-}$ with $|a^{\hat r}|$ peaks (dots) inside; the subcritical curve diverges.}
\end{figure}

\subsubsection{Polar repulsion}
At the outer shell $r_{+}=\mmj/2$, $\sqrt\Delta\to 0$ and $\cG\to 0$, while the Boyer--Lindquist offset $\rBL-{\rBL}_{+}\propto\Delta$ vanishes faster than $\sqrt\Delta$ itself.  The leading divergence exponents of the Ricci and Kretschmann scalars are angle-dependent.  Off-axis ($\sin\vartheta\neq 0$), we have
\be
\ba{lll}
R\sim(r-r_{+})^{-2\alpha_{\zeta}}\,,\quad&\quad K\sim(r-r_{+})^{-4\alpha_{\zeta}}\,,\quad&\quad
\alpha_{\zeta}=\dfrac{\mmj^{2}+\sigma(\zeta)\,\mmj|\qq|+\qq^{2}}{\mmj^{2}}\,.
\ea
\label{rPlusDiv_body}
\ee
Here $\sigma(0)=\mathrm{sgn}\,\qq$ on the FJNW dilaton branch, while $\sigma(\zeta)=-1$ for $0<\zeta<\pi/2$ on branches with axion.  The exponent $\alpha_{\zeta}$ is positive for every ${\qq\neq 0}$ and every $\zeta$.  On the rotation axis ($\sin\vartheta=0$) the angular factor evaluates to unity, the $\qq^{2}/\mmj^{2}$ contribution drops, and the exponent reduces to $\alpha^{\rm axis}_{\zeta}=1+\sigma(\zeta)|\qq|/\mmj$, which becomes negative in the polar repulsion regime~(\ref{repCrit}) so $R,K\to 0$ on axis at the shell (appendix).  The outer shell is a Riemannian curvature singularity at the Kerr horizon position, angularly concentrated and strongest off axis, except on the rotation axis in the supercritical regime where it becomes regular in curvature.  On the supercritical rotation axis the \emph{inverse} metric $g^{\mu\nu}$ stays finite, while the lower-index Riemannian metric components diverge; off axis the inverse components can themselves diverge (appendix).

Closed-string probes couple to the string frame metric at leading order in $\alpha'$ and therefore follow string frame geodesics; in this frame the dilaton kinetic term enters with the opposite sign to a canonical scalar, permitting outward scalar-induced acceleration.  The same sign structure has also been used in closed-string models of late-time cosmic acceleration~\cite{Lee:2023boi}.  The orthonormal proper acceleration of a static observer, $a^{\hat r}=\sqrt{g_{rr}}\,a^{r}$, with $a^{\mu}=u^{\nu}\nabla_{\nu}u^{\mu}$ and $u^{\mu}=\delta^{\mu}_{t}/\sqrt{-g_{tt}}$, evaluates to
\be
a^{\hat r}=\dfrac{e^{-\phi}\sqrt\Delta}{r\,\sqrt{\cB}\,\Sigma_{+}\,\Lambda}\,\cF\,,\quad
\cF(r,\vartheta)\!\equiv\!\mm\Sigma_{-}+\dfrac{r\Sigma_{+}\Lambda}{\sqrt\Delta}\partial_{r}\phi\,,
\label{Frep}
\ee
with strictly positive prefactor in the stationary region, so $\mathrm{sign}\,a^{\hat r}=\mathrm{sign}\,\cF$.  Fig.~\ref{fig:polar} displays the corresponding repulsive zones and axial acceleration profiles.  At the intersection of the rotation axis and the outer shell on branches with axion ($\zeta\in(0,\pi/2)$), $\cF_{\rm pole}(r_{+})=2\mm(\mm+\mmj)(\mmj-|\qq|)$; a polar geodesic released near $r_{+}$ accelerates outward if and only if
\be
\boxed{\;|\qq|>\mmj\;}\,.
\label{repCrit}
\ee
This is the rotation-dependent threshold.  Above this threshold, dilaton-driven barriers in the axial timelike and null effective potentials shield $r_{+}$ from axial probe geodesics; see the appendix.  The mechanism is the dilaton's negative kinetic term in string frame: it is invisible to a curvature diagnostic alone but controls the sign of the string frame acceleration.

\begin{figure}[h]
\centering
\includegraphics[width=0.6\linewidth]{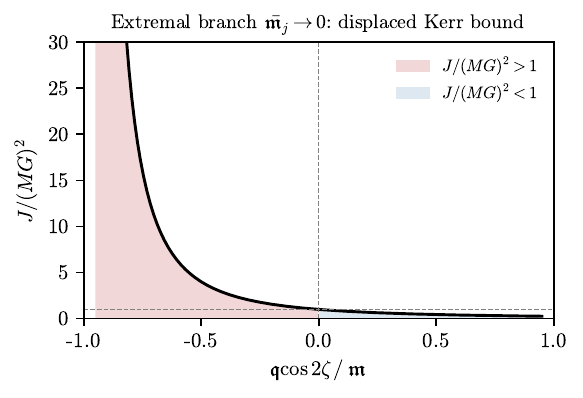}
\caption{\label{fig:s1ext}Displaced Kerr bound on the positive extremal branch $j\to\mm$: $J/(MG)^{2}=\mm^{2}/(\mm+\qq\cos 2\zeta)^{2}$, reducing to the Kerr bound $J/(MG)^{2}=1$ only when $\qq\cos 2\zeta=0$.  Negative $\qq\cos 2\zeta$ permits $J/(MG)^{2}>1$ (red shaded), in contrast with Kerr.}
\end{figure}

On the positive extremal branch $j\to\mm$ (so $\mmj\to 0$), the factors with exponents $\pm 2\qq/\mmj$ in~(\ref{dilaton}) have the smooth limit $e^{2\phi}\to\cos^{2}\!\zeta\,e^{-2\qq/r}+\sin^{2}\!\zeta\,e^{2\qq/r}$; the threshold $\mmj$ tends to zero, and $J=\mm^{2}$ together with~(\ref{MGbetagamma}) gives $J/(MG)^{2}=\mm^{2}/(\mm+\qq\cos 2\zeta)^{2}$, displaced from the Kerr bound by the dilaton charge $\qq\cos 2\zeta$ (negative $\qq\cos 2\zeta$ permits $J/(MG)^{2}>1$, Fig.~\ref{fig:s1ext}).  For $j\to-\mm$, the same statement applies to $|J|/(MG)^{2}$.

\vspace{5pt}

\section{Discussion}
\label{sec:discussion}

\subsection{Relation to the EMDA Killing-tensor families}
\label{sec:EMDA-detailed}

In the GGK/GK Killing-tensor families~\cite{Garcia:1995qz,Galtsov:2025nia}, the complex axidilaton charge $\mathcal{D}$ is algebraically locked to the complex electromagnetic charge $\mathcal{Q}$,
\be
\mathcal{D}\;=\;-\,\frac{\mathcal{Q}^{\ast\,2}}{2\mathcal{M}}\,,
\label{GKdilatonlock}
\ee
($\mathcal{M}$ the complex mass; their Eq.~(80)), and the axidilaton $\lambda=\chi+ie^{-2\phi}$ is fractional linear (a M\"obius transformation); when the Maxwell charge is set to $\mathcal{Q}=0$, the scalar charges therefore vanish.  By contrast, our solution keeps the scalar charge $\qq$ free at $\mathcal{Q}=0$, and the dilaton~(\ref{dilaton}) is a fractional power of $\cG(r)$ with noninteger exponent $2\qq/\mmj$, placing it outside this Killing tensor class.  The two families meet only at $\qq=0$, pure Kerr.  In contrast to EMDA, the scalar charge $\qq$ remains an independent hair even in the absence of a Maxwell sector ($\mathcal{Q}=0$).

\subsection{Non-Riemannian interpretation}
\label{sec:nonriem}
The supercritical polar shell ($|\qq|>\mmj$) carries an unusual metric degeneracy: on the rotation axis the inverse metric $g^{\mu\nu}$ stays finite, while the lower-index Riemannian metric components diverge.  This suggests, but does not establish, a non-Riemannian DFT interpretation~\cite{Morand:2021}, since the fundamental variables are $\{d,\cH_{AB}\}$ rather than $g_{\mu\nu}$ alone.  We do not construct an explicit non-Riemannian generalized metric for the shell here; a full non-Riemannian completion of its geometry and an analysis of geodesic completeness remain open problems.

\subsection{Fuzzball-like degeneracy and outlook}
\label{sec:fuzzball-outlook}
The first structural result is the geometric memory of rotation.  In pure GR the Kerr family returns to Schwarzschild at $j=0$, but the pure NS--NS rotating branch returns instead to a closed-string-dressed ZV geometry carrying $H$-flux.  The scalar hair therefore leaves an oblate spatial imprint even after the angular momentum is removed.  This is the sharpest distinction between the present vacuum and the Maxwell-supported Kerr--Sen/EMDA families: the latter use gauge charge to keep the static limit spherical, whereas the pure NS--NS construction stores the hair in the spatial multipoles.

The second result is monopole degeneracy.  The axial branch and the spherical BMQ branch have the same $\ell=0$ data $\{MG,\betappn,\gammappn,h\}$, so asymptotic PPN measurements do not determine the bulk geometry.  The corresponding BMQ reality bound maps to $4\qq^{2}\geq h^{2}$ and is automatic for $h=-2\qq\sin2\zeta$; $h=0$ is only the FJNW/dilaton-only slice.  The degeneracy is lifted at $\ell=2$, where the ZV branch carrying $H$-flux has higher multipoles while spherical BMQ does not.  At minimum, within the presently known solution space,
\be
\Omega(MG,\qq,h)\geq 2\,,
\ee
suggesting a classical degeneracy of geometries at fixed monopole data, in the spirit of fuzzball constructions~\cite{Mathur:2005zp}, without yet supplying a systematic state count.

The third result is dynamical.  Rotation produces a stringy Lense--Thirring correction controlled by $\qq\cos2\zeta$ and a strong-field polar repulsion threshold $|\qq|>\mmj$.  The latter is the rotating closed-string uplift of the FJNW-type axial repulsion of the static dilaton branch: in the supercritical regime the axis becomes regular in curvature at the shell, axial probes are repelled before reaching it, and the peak axial acceleration is of order $\mathcal{O}(\mm^{-1})$, invisible to the asymptotic PPN expansion.

Finally, the same supercritical shell is where a non-Riemannian DFT interpretation may become relevant.  What we have established is the kinematic input: on the rotation axis the inverse metric remains finite, while the lower-index Riemannian metric components diverge, and the fundamental DFT variables are $\{d,\cH_{AB}\}$ rather than $g_{\mu\nu}$ alone.  The next problems are therefore to classify the static axisymmetric moduli at fixed $\{MG,\qq,h\}$, construct the full non-Riemannian DFT completion of the outer shell $r_{+}$~\cite{Morand:2021}, and determine how the polar geodesic shielding extends beyond the axial probes treated here.\\

\noindent\textbf{Code availability.} A Maple/GRTensorIII worksheet and an independent Mathematica notebook used for analytic verification are publicly archived at~\cite{Code:GitHub} under the MIT License.  Portions of the code and the figures were prepared with the assistance of ChatGPT and Claude.\vspace{5pt}

\noindent\textbf{Acknowledgments.} This work is supported by the National Research Foundation of Korea (NRF) through grants RS-2025-25414114, RS-2023-NR077094, and RS-2020-NR049598 (Center for Quantum Spacetime).\\

\newpage

%======================== BIBLIOGRAPHY ========================

\appendix
\begin{center}
\Large{\bf{APPENDIX}}
\end{center}

\section{Derivations}

\subsection{Derivation of the $\ODD$-singlet dilaton density~(\ref{DFTdilaton})}
The string frame line element~(\ref{metric}) is block diagonal in $(t,\varphi)\!\oplus\!(r,\vartheta)$, so the metric determinant factorizes as
\be
\det(g_{\mu\nu})=g_{rr}\,g_{\vartheta\vartheta}\,\big(g_{tt}\,g_{\varphi\varphi}-g_{t\varphi}^{\,2}\big)\,.
\label{eq:detBlock}
\ee
The diagonal $(r,\vartheta)$ block gives $g_{rr}g_{\vartheta\vartheta}=e^{4\phi}\cB^{2}r^{2}$.  In the $(t,\varphi)$ block, the $\omega^{2}$ terms in $g_{tt}g_{\varphi\varphi}$ and $g_{t\varphi}^{\,2}$ cancel exactly, leaving $g_{tt}g_{\varphi\varphi}-g_{t\varphi}^{\,2}=-e^{4\phi}\Delta\sin^{2}\!\vartheta$.  Hence
\be
g\equiv\det(g_{\mu\nu})=-e^{8\phi}\cB^{2}r^{2}\Delta\sin^{2}\!\vartheta\,,\quad\sqrt{-g}=e^{4\phi}\cB\,r\sqrt{\Delta}\,\sin\vartheta\,,
\label{eq:sqrtg}
\ee
so that $e^{-2d}=e^{-2\phi}\sqrt{-g}=e^{2\phi}\cB\,r\sqrt{\Delta}\,\sin\vartheta$; substituting~(\ref{Bomega}) for $\cB$ yields~(\ref{DFTdilaton}).

\subsection{The solution in Boyer--Lindquist coordinates}
\label{app:BL}
\renewcommand{\rBL}{\bar{r}}%
For brevity, we write $\bar r$ for the Boyer--Lindquist coordinate throughout the remainder of this appendix; an unbarred $r$ denotes the original quasi-isotropic coordinate only when the coordinate map or the shell position $r_{+}$ is being discussed.  The boxed configuration~(\ref{dilaton}),~(\ref{Hflux}),~(\ref{metric}) admits a closed-form rewriting in Boyer--Lindquist coordinates $(t,\rBL,\vartheta,\varphi)$ via $\rBL(r)=r+\mm+\mmj^{2}/(4r)$.  Introduce the inner and outer shells $\rBL_{\pm}\equiv\mm\pm\mmj$ (so $\rBL_{+}\rBL_{-}=j^{2}$), giving
\be
\Delta=(\rBL-\rBL_{+})(\rBL-\rBL_{-})\,,\qquad
\Sigma_{+}=\rBL^{2}+j^{2}\cos^{2}\!\vartheta\,.
\label{eq:DeltaSigmaBL}
\ee
Thus the quasi-isotropic $\Delta=(r-\mmj^{2}/4r)^{2}$ becomes the usual Boyer--Lindquist polynomial; no new function is being introduced.  Using the perfect square identity $\cG(r)^{2\qq/\mmj}=[(\rBL-\rBL_{+})/(\rBL-\rBL_{-})]^{\qq/\mmj}$ and the chain rule $\rd r=(r/\sqrt{\Delta})\,\rd\rBL$, the quasi-isotropic conformal warp block rewrites as $\cB(\rd r^{2}+r^{2}\rd\vartheta^{2})=\Sigma_{+}\,(1+\mmj^{2}\sin^{2}\!\vartheta/\Delta)^{-\qq^{2}/\mmj^{2}}(\rd\rBL^{2}/\Delta+\rd\vartheta^{2})$.  The dilaton, Kalb--Ramond three-form, and string frame line element then read, respectively,
\be
\boxed{\;e^{2\phi(\rBL)}=\left(\dfrac{\rBL-\rBL_{+}}{\rBL-\rBL_{-}}\right)^{\!\qq/\mmj}\!\cos^{2}\!\zeta+\left(\dfrac{\rBL-\rBL_{+}}{\rBL-\rBL_{-}}\right)^{\!-\qq/\mmj}\!\sin^{2}\!\zeta\,,\;}
\label{eq:dilatonBL}
\ee
\be
\boxed{\;H_{(3)}=h\sin\vartheta\,\rd t\wedge\rd\vartheta\wedge\rd\varphi\,,\quad h=-2\qq\sin 2\zeta\,,\;}
\label{eq:HfluxBL}
\ee
\be
\boxed{\;\rd s^{2}=e^{2\phi(\rBL)}\!\left[\scalebox{0.88}{$-\dfrac{\Lambda}{\Sigma_{+}}\big(\rd t-\omega(\rBL,\vartheta)\rd\varphi\big)^{2}+\Sigma_{+}\!\left(1+\dfrac{\mmj^{2}\sin^{2}\!\vartheta}{\Delta}\right)^{\!-\qq^{2}/\mmj^{2}}\!\!\left(\dfrac{\rd\rBL^{2}}{\Delta}+\rd\vartheta^{2}\right)+\dfrac{\Sigma_{+}\Delta\sin^{2}\!\vartheta}{\Lambda}\rd\varphi^{2}$}\right]\,,\;}
\label{eq:metricBL}
\ee
with frame-dragging function $\omega(\rBL,\vartheta)=-2\mm j\,\rBL\sin^{2}\!\vartheta/\Lambda$ and local Kalb--Ramond potential $B_{t\varphi}=h\cos\vartheta$.  The outer shell $r_{+}=\mmj/2$ maps to $\rBL_{+}=\mm+\mmj$, the standard Kerr horizon locus; at $\qq=0$ the bracket in~(\ref{eq:metricBL}) reduces to the Boyer--Lindquist Kerr line element.  Eqs.~(\ref{eq:dilatonBL})--(\ref{eq:metricBL}) are verified to solve~(\ref{EDFE}) using a Maple/GRTensorIII worksheet and an independent Mathematica notebook archived at~\cite{Code:GitHub}; both perform the verification natively in Boyer--Lindquist coordinates.

\subsection{Coordinate-invariant Ricci scalar diagnostics}
At $j=0$ the string frame Ricci scalar reads
\be
R\big|_{j=0}=-\Big(2e^{-2\phi}+3e^{-6\phi}\sin^{2}\!2\zeta\Big)R^{E}\big|_{j=0}\,,\qquad
R^{E}\big|_{j=0}=\dfrac{2\qq^{2}}{\rBL^{2}\Delta}\!\left(1+\dfrac{\mm^{2}\sin^{2}\!\vartheta}{\Delta}\right)^{\!\qq^{2}/\mm^{2}}\!.
\label{eq:Ricci}
\ee
Both scalars are $\vartheta$-dependent for $\qq\neq 0$ and vanish at $\qq=0$ (Schwarzschild).  The $j\to 0$ slice is therefore a Zipoy--Voorhees geometry dressed by closed-string fields, not the spherical BMQ branch.

\subsection{Curvature divergence exponents at $r_{+}$}
Set $\bar\epsilon\equiv r-r_{+}\to 0^{+}$, so that $\sqrt\Delta\approx 2\bar\epsilon$ and the Boyer--Lindquist offset $\rBL-{\rBL}_{+}\approx 2\bar\epsilon^{2}/\mmj$.  The leading divergence exponents at the outer shell depend on angle, so the cases away from the axis and on the axis require separate formulas.

\textit{Off-axis} ($\sin\vartheta\neq 0$).  The Ricci and Kretschmann scalars diverge as
\be
R\sim\bar\epsilon^{\,-2\alpha_{\zeta}}\,,\quad
K\sim\bar\epsilon^{\,-4\alpha_{\zeta}}\,,\quad
\alpha_{\zeta}\equiv\dfrac{\mmj^{2}+\sigma(\zeta)\,\mmj|\qq|+\qq^{2}}{\mmj^{2}}\,,
\label{eq:rPlusDivRot}
\ee
with $\sigma(\zeta\!=\!0)=\mathrm{sgn}\,\qq$ on the FJNW dilaton branch and $\sigma(\zeta\!\neq\!0)=-1$ on branches with axion; the FJNW sign tracks the dilaton orientation, while the axion branch is symmetric under $\qq\to-\qq$.  The static limit $j\to 0$ recovers $\mmj\to\mm$.  The exponent is positive for every $\qq\neq 0$ and every $\zeta$, with the divergence weakening monotonically from the equator toward the poles through the angular factor $(1+\mmj^{2}\sin^{2}\!\vartheta/\Delta)^{-\qq^{2}/\mmj^{2}}$.  A numerical log-log fit at $\mm=1$, $j=3/5$, $\qq=1/2$, $\zeta=0$, $\vartheta=\pi/3$ confirms the predicted exponent $\alpha\simeq 2.0156$ with relative error below $10^{-3}$.

\textit{On-axis} ($\sin\vartheta=0$).  The angular factor evaluates to unity identically, removing the $\qq^{2}/\mmj^{2}$ contribution, so
\be
R\sim\bar\epsilon^{\,-2\alpha^{\rm axis}_{\zeta}}\,,\quad
K\sim\bar\epsilon^{\,-4\alpha^{\rm axis}_{\zeta}}\,,\quad
\alpha^{\rm axis}_{\zeta}\equiv 1+\sigma(\zeta)\,|\qq|/\mmj\,<\,\alpha_{\zeta}\,.
\label{eq:rPlusDivAxis}
\ee
In the polar repulsion regime ($|\qq|>\mmj$ on branches with axion, $\qq<-\mmj$ on FJNW), $\alpha^{\rm axis}_{\zeta}<0$ and $R,K\to 0$ at the shell on the axis: the axis is regular in curvature at $r_{+}$ precisely when the polar repulsion turns on.

The outer shell is therefore a Riemannian curvature singularity at the Kerr horizon position, angularly concentrated and strongest at the equator (weakening to the poles), except on the rotation axis in the supercritical regime where it becomes regular in curvature.

\subsection{Inverse-metric scaling near the shell}
Whether the inverse string frame metric stays finite at the shell is a question local to the axis.  Write $\bar\epsilon\equiv r-r_{+}\to 0^{+}$ and $a\equiv|\qq|/\mmj$; on branches with axion $e^{2\phi}\sim\bar\epsilon^{\,-2a}$.  Away from the axis ($\sin\vartheta\neq 0$) the angular factor in $\cB$ scales as $(1+\mmj^{2}\sin^{2}\!\vartheta/\Delta)^{-\qq^{2}/\mmj^{2}}\sim\bar\epsilon^{\,2a^{2}}$, so that
\be
g^{rr}=e^{-2\phi}\cB^{-1}\sim\bar\epsilon^{\,2a(1-a)}\,,\qquad
g^{tt},\,g^{t\varphi},\,g^{\varphi\varphi}\sim\bar\epsilon^{\,2a-2}\,;
\ee
the radial component diverges for $a>1$ and the $(t,\varphi)$ block for $a<1$, so away from the axis the inverse metric is never globally finite.  On the rotation axis ($\sin\vartheta=0$, $\omega=0$, $\Lambda=\Delta$) the angular factor is unity and
\be
g^{rr}\big|_{\vartheta=0}\sim\bar\epsilon^{\,2a}\,,\qquad
g^{tt}\big|_{\vartheta=0}=-e^{-2\phi}\Sigma_{+}/\Delta\sim-\bar\epsilon^{\,2a-2}\,.
\ee
Both displayed inverse components are finite for $a>1$.  Thus only in the supercritical regime $a>1$, and only on the axis, does the inverse metric stay finite while the lower-index Riemannian metric components diverge.  This is the precise sense, local to the axis, in which the polar shell is a candidate for a non-Riemannian DFT description.

\subsection{Ergosurface structure}
The Killing vector $\partial_{t}$ has norm $g_{tt}=-e^{2\phi}\Lambda/\Sigma_{+}$.  Since $e^{2\phi}>0$, the stationary limit surface is determined by $\Lambda=0$, giving
\be
{\rBL}_{\rm ergo}(\vartheta)=\mm+\sqrt{\mm^{2}-j^{2}\cos^{2}\!\vartheta}\,,
\label{eq:ergoSurface}
\ee
identical to Kerr: the closed-string hair enters $g_{tt}$ only through the strictly positive prefactor $e^{2\phi}$ and does not displace the ergosurface.  The radial separation from the ${\rBL}_{+}$ shell,
\be
\Delta_{\rm ergo}(\vartheta)=\sqrt{\mm^{2}-j^{2}\cos^{2}\!\vartheta}-\mmj\,\geq\,0\,,
\label{eq:ergoThickness}
\ee
vanishes on the rotation axis (where the ergosurface intersects the outer shell) and attains its maximum $\mm-\mmj$ at the equator.  Closed-string hair preserves the Kerr ergoregion exactly; only the horizon-side edge of the ergoregion changes from a regular event horizon ($\qq=0$) to the locus where the horizon and curvature singularity coincide.

\subsection{Derivation of the first integral~(\ref{eq:firstInt_body})}
Passing to the Einstein frame $g^{E}_{\mu\nu}=e^{-2\phi}g_{\mu\nu}$ and dualizing the $B$-field to a four-dimensional axion scalar $\chi$ via ${\star}_{E}(e^{-4\phi}H_{(3)})=\rd\chi$, the NS--NS vacuum equations~(\ref{EDFE}) reduce to the Einstein--dilaton--axion system with dilaton equation of motion $\Box_{E}\phi=\tfrac{1}{2}e^{4\phi}(\partial\chi)^{2}$.  In the Boyer--Lindquist chart, for purely radial $\phi=\phi(\rBL)$, the $\vartheta$ dependence of $\sqrt{-g_{E}}g_{E}^{\rBL\rBL}=\Delta\sin\vartheta$ trivializes; the Hodge dual axion derivative reads $\partial_{\rBL}\chi=-he^{-4\phi}/\Delta$, and the radial dilaton equation becomes
\be
\partial_{\rBL}(\Delta\,\partial_{\rBL}\phi)=\dfrac{h^{2}\,e^{-4\phi}}{2\Delta}\,.
\label{eq:radODE_E}
\ee
Multiplying by $4\Delta\,\partial_{\rBL}\phi$ and using $e^{-4\phi}\partial_{\rBL}\phi=-\tfrac{1}{4}\partial_{\rBL}e^{-4\phi}$ converts both sides into total $\rBL$ derivatives, yielding, upon integration, the main text equation~(\ref{eq:firstInt_body}), valid throughout the exterior $\rBL>{\rBL}_{+}$.  Asymptotic consistency provides a check: at spatial infinity $\cG\to 1$ gives $\Delta\,\partial_{\rBL}\phi\to\qq\cos 2\zeta$ and $\phi\to 0$, so $4\qq^{2}\cos^{2}\!2\zeta+h^{2}=4\qq^{2}$ recovers exactly $h=-2\qq\sin 2\zeta$.  Equation~(\ref{eq:firstInt_body}) is the rotating generalization of the static first integral [Eq.~(4.56) of~\cite{Angus:2018mep}; Eq.~(B.23) of~\cite{Ko:2016dxa}]; combined with the surface integral monopole dictionary, the $\SO(2)_{\zeta}$-invariant value $4\qq^{2}$ ties the four parameters $(\mm,j,\qq,\zeta)$ to the asymptotic monopole charges through a nonlinear conservation law consistent with DFT.

\subsection{Derivation of the PPN form~(\ref{PPN}) and asymptotic charges~(\ref{MGbetagamma})}
The quasi-isotropic large-$r$ expansion determines the four asymptotic charges.  Since $\ln\cG=-\mmj/r+\mathcal{O}(r^{-3})$, one has $\cG^{\pm 2\qq/\mmj}=1\mp 2\qq/r+2\qq^{2}/r^{2}+\mathcal{O}(r^{-3})$, so the dilaton~(\ref{dilaton}) gives
\be
e^{2\phi}=1-\dfrac{2\qq\cos 2\zeta}{r}+\dfrac{2\qq^{2}}{r^{2}}+\mathcal{O}(r^{-3})\,.
\ee
From $\Delta=r^{2}-\mmj^{2}/2+\mathcal{O}(r^{-2})$ and $\Sigma_{+}=r^{2}+2\mm r+(\mm^{2}+\mmj^{2}/2+j^{2}\cos^{2}\!\vartheta)+\mathcal{O}(r^{-1})$, the ratio $\Lambda/\Sigma_{+}=1-2\mm/r+2\mm^{2}/r^{2}+\mathcal{O}(r^{-3})$ is angle independent at this order.  Multiplying by $e^{2\phi}$,
\be
-g_{tt}=1-\dfrac{2(\mm+\qq\cos 2\zeta)}{r}+\dfrac{2(\mm^{2}+2\mm\qq\cos 2\zeta+\qq^{2})}{r^{2}}+\mathcal{O}(r^{-3})\,.
\ee
Matching~(\ref{PPN}) fixes $MG=\mm+\qq\cos 2\zeta$ and $\betappn(MG)^{2}=(MG)^{2}+\qq^{2}\sin^{2}\!2\zeta$, hence $\betappn=1+\qq^{2}\sin^{2}\!2\zeta/(MG)^{2}$.

For $g_{rr}=e^{2\phi}\cB$ with $\cB=(\Sigma_{+}/r^{2})(1+\mmj^{2}\sin^{2}\!\vartheta/\Delta)^{-\qq^{2}/\mmj^{2}}$, the two factors expand as
\be
\frac{\Sigma_{+}}{r^{2}}=1+\frac{2\mm}{r}+\frac{\mm^{2}+\mmj^{2}/2+j^{2}\cos^{2}\!\vartheta}{r^{2}}+\mathcal{O}(r^{-3})\,,
\label{eq:SigmaExp}
\ee
\be
\left(1+\frac{\mmj^{2}\sin^{2}\!\vartheta}{\Delta}\right)^{-\qq^{2}/\mmj^{2}}=1-\frac{\qq^{2}\sin^{2}\!\vartheta}{r^{2}}+\mathcal{O}(r^{-4})\,,
\label{eq:HairExp}
\ee
The second expansion uses $\Delta=r^{2}-\mmj^{2}/2+\mathcal{O}(r^{-2})$.  Multiplying the two factors gives
\be
\cB(r,\vartheta)=1+\frac{2\mm}{r}+\frac{\mm^{2}+\mmj^{2}/2+j^{2}\cos^{2}\!\vartheta-\qq^{2}\sin^{2}\!\vartheta}{r^{2}}+\mathcal{O}(r^{-3})\,.
\label{eq:Bexp}
\ee
The $1/r$ coefficient $2\mm$ is angle independent, accounting for the leading monopole degeneracy between the ZV branch carrying $H$-flux and the spherical BMQ branch; the $1/r^{2}$ coefficient carries the $\vartheta$ dependence $j^{2}\cos^{2}\vartheta-\qq^{2}\sin^{2}\vartheta$ that distinguishes the two branches at $\ell=2$.  Multiplying~(\ref{eq:Bexp}) by $e^{2\phi}$ gives $g_{rr}=1+2(\mm-\qq\cos 2\zeta)/r+\mathcal{O}(r^{-2})$, and hence $\gammappn=(\mm-\qq\cos 2\zeta)/(\mm+\qq\cos 2\zeta)$.  Equivalently $\gammappn-1=-2\qq\cos 2\zeta/MG$, consistent with $e^{2\phi}\simeq 1+(\gammappn-1)MG/r$ in~(\ref{PPN}).

Finally, with $\omega=-2\mm j\,\rBL\sin^{2}\!\vartheta/\Lambda$ the cross-term reads $g_{t\varphi}=-e^{2\phi}\,2\mm j\,\rBL\sin^{2}\!\vartheta/\Sigma_{+}\simeq -2\mm j\sin^{2}\!\vartheta/r$, so $J=\mm j$.  Thus the asymptotic charges are
\be
\boxed{\,MG=\mm+\qq\cos 2\zeta\,,\quad\; \betappn=1+\dfrac{\qq^{2}\sin^{2}\!2\zeta}{(MG)^{2}}\,,\quad\; \gammappn=\dfrac{\mm-\qq\cos 2\zeta}{\mm+\qq\cos 2\zeta}\,,\quad\; J=\mm j\,}\,.
\ee
The former three quantities are $j$-independent at leading order and coincide with the static BMQ results of Ref.~\cite{Choi:2022fqj}, while $J=\mm j$ is the rotational charge.  The Cassini Shapiro-delay bound $|\gammappn-1|\lesssim 2.3\times 10^{-5}$~\cite{Bertotti:2003rm} constrains $|\qq\cos 2\zeta|/MG\lesssim 1.15\times 10^{-5}$, as for the Sun.

\subsection{Komar-type surface-integral charges}
Each of the four parameters is also realized as an invariant boundary charge at spatial infinity, from integrals built on the asymptotic Killing vectors $\partial_{t}$ and $\partial_{\varphi}$,
\begin{align}
MG\;&=\;-\frac{1}{8\pi}\oint_{\infty}\!\rd S_{\mu\nu}\,e^{-2\phi}\,\nabla^{\mu}\xi^{\nu}_{(t)}\,,\label{Mint}\\
J\;&=\;\frac{1}{16\pi}\oint_{\infty}\!\rd S_{\mu\nu}\,e^{-2\phi}\,\nabla^{\mu}\xi^{\nu}_{(\varphi)}\,,\qquad j=\frac{J}{\mm}\,,\label{Jint}\\
\qq\cos 2\zeta\;&=\;-\frac{1}{8\pi}\oint_{\infty}\!\rd S_{i}\,\partial^{i}e^{-2\phi}\,,\label{Qcosint}\\
h\;&=\;\frac{1}{8\pi}\oint_{\infty}\!\rd S_{i}\,\epsilon^{ijk}H_{tjk}\;=\;-2\qq\sin 2\zeta\,.\label{hconst}
\end{align}
Here $\xi_{(t)}=\partial_{t}$, $\xi_{(\varphi)}=\partial_{\varphi}$, and the $e^{-2\phi}$ weight is adapted to the DFT dilaton measure $e^{-2d}=e^{-2\phi}\sqrt{-g}$.  Inserting the large-$r$ expansion of the preceding paragraph into~(\ref{Mint})--(\ref{Jint}) reproduces $MG=\mm+\qq\cos 2\zeta$ and $J=\mm j$, while~(\ref{Qcosint})--(\ref{hconst}) read off the $1/r$ falloffs of $\phi$ and $B_{t\varphi}=h\cos\vartheta$.  This is the same electric $H$-flux parameter as in the static BMQ vacuum family, where the reality condition is $b^{2}\geq h^{2}$ with $b$ the BMQ scalar-charge parameter~\cite{Ko:2016dxa}; in the present parametrization $b^{2}=4\qq^{2}$ and $h=-2\qq\sin2\zeta$, so the bound is automatic.  Together with the first integral~(\ref{eq:firstInt_body}), the scalar coset charge vector is $(\qq\cos 2\zeta,-h/2)=\qq(\cos 2\zeta,\sin 2\zeta)$.  Its invariant magnitude and phase therefore separate algebraically as $|\qq|=\sqrt{(\qq\cos 2\zeta)^{2}+(h/2)^{2}}$ and, when $\qq\cos 2\zeta\neq0$, $\tan 2\zeta=-h/\bigl[2(\qq\cos 2\zeta)\bigr]$; at the boundary loci the phase is read by continuity.  The sign of $\qq$ is the orientation of this charge vector, not the positive square root.  With the surface integral dictionary~(\ref{Mint})--(\ref{hconst}) this furnishes the complete DFT identification of the four parameters, the rotating analogue of the static shell theorem~\cite{Ko:2016dxa}.

\subsection{Explicit form of $\cF$}
\begin{figure}[h]
\centering
\includegraphics[width=0.7\linewidth]{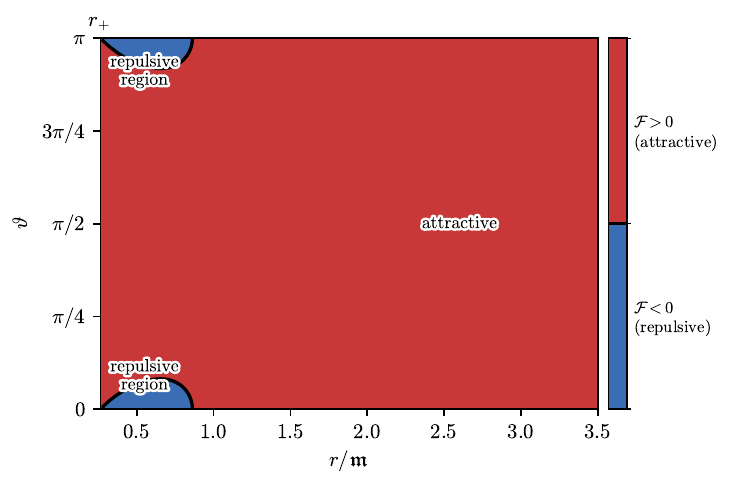}
\caption{\label{fig:polar2d}Sign of $\cF(r,\vartheta)$ over $(r/\mm,\vartheta)$ at $\mm=1$, $j=0.85$, $\zeta=0.5$, $|\qq|/\mmj=1.5$: blue marks the repulsive zones ($\cF<0$) and red the attractive bulk ($\cF>0$).  The black $\cF=0$ contour bounds the $\vartheta$-localized repulsive regions wrapping the rotation axis $\vartheta=0,\pi$.}
\end{figure}

Differentiating the dilaton~(\ref{dilaton}) and using the quasi-isotropic identity $\partial_{r}\cG/\cG=\mmj/(r\sqrt\Delta)$, one obtains
\be
\sqrt\Delta\,\partial_{r}\phi=\frac{\qq\,(\cG^{2\qq/\mmj}\cos^{2}\!\zeta-\cG^{-2\qq/\mmj}\sin^{2}\!\zeta)}{r\,e^{2\phi}}\,.
\label{eq:sqrtDeltaphi}
\ee
Substituting into the main text expression~(\ref{Frep}) gives the explicit closed-form expression
\be
\cF(r,\vartheta)=\mm\Sigma_{-}+\frac{\qq\,\Sigma_{+}\Lambda\,(\cG^{2\qq/\mmj}\cos^{2}\!\zeta-\cG^{-2\qq/\mmj}\sin^{2}\!\zeta)}{\Delta\,e^{2\phi}}\,,
\label{eq:Fexplicit}
\ee
valid throughout the exterior $r>r_{+}$ for all $\vartheta$, $\zeta\in[0,\pi/2)$, and $\qq\in\mathbb{R}$; its sign over $(r,\vartheta)$ is mapped in Fig.~\ref{fig:polar2d}.

\textit{Pole limit.}  At $\vartheta=0$, $r=r_{+}=\mmj/2$ one has $\Sigma_{+}|_{r_{+},\vartheta=0}=2\mm\bar r_{+}$, $\bar r_{+}^{2}-j^{2}=2\mmj\bar r_{+}$ with $\bar r_{+}\equiv\bar r(r_{+})=\mm+\mmj$.  On branches with axion ($0<\zeta<\pi/2$) the regularized limit reads $\sqrt\Delta\,\partial_{r}\phi|_{r_{+}}=-2|\qq|/\mmj$, giving $\cF_{\rm pole}(r_{+})=2\mm\bar r_{+}(\mmj-|\qq|)$ and the symmetric criterion $|\qq|>\mmj$ used in the main text~(\ref{Frep})--(\ref{repCrit}).  On the FJNW dilaton branch ($\zeta=0$) the dilaton has only one factor $e^{2\phi}=\cG^{2\qq/\mmj}$, yielding $\sqrt\Delta\,\partial_{r}\phi|_{r_{+}}=2\qq/\mmj$ and $\cF_{\rm pole}(r_{+})=2\mm\bar r_{+}(\mmj+\qq)$; the repulsion criterion is one-sided, $\qq<-\mmj$, since only negative-$\qq$ FJNW dilatons diverge upward at $r_{+}$.

\subsection{Timelike geodesic turning-point derivation of the polar criterion}
\begin{figure}[h]
\centering
\includegraphics[width=0.6\linewidth]{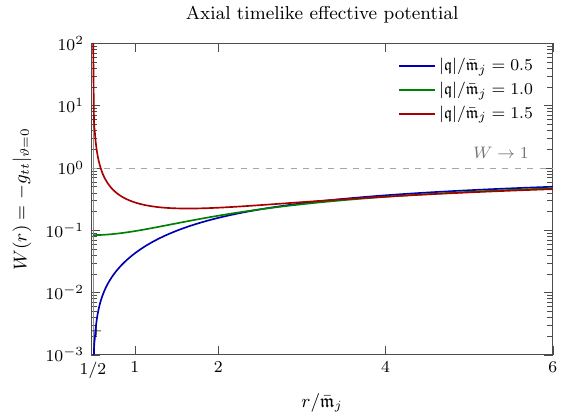}
\caption{\label{fig:effpot}Axial timelike effective potential $W(r)=-g_{tt}|_{\vartheta=0}$ at $\mm=1$, $j=0.85$, $\zeta=0.5$, plotted on a logarithmic scale for $|\qq|/\mmj=0.5,1.0,1.5$.  The thin vertical line marks $r_{+}/\mmj=1/2$.  In accord with the threshold~(\ref{repCrit}), the $|\qq|/\mmj=0.5$ subcritical curve falls to zero, the $|\qq|/\mmj=1.0$ critical curve tends to a finite limit, and the $|\qq|/\mmj=1.5$ supercritical curve diverges, forming the inner barrier.}
\end{figure}

The main text criterion $\cF<0$ follows independently from the polar timelike geodesic equation, providing a kinematic confirmation independent of the static observer construction.  On the rotation axis ($\vartheta=0$), with vanishing conserved azimuthal angular momentum $L\equiv p_{\varphi}=0$, one has $\omega=0$, $\Lambda=\Delta$, and $\cB|_{\vartheta=0}=\Sigma_{+}/r^{2}$.  A timelike worldline restricted to $(t,r)$ obeys $g_{tt}\dot t^{2}+g_{rr}\dot r^{2}=-1$ with stationary conserved energy $E=-g_{tt}\dot t$.  Eliminating $\dot t$,
\be
\dot r^{2}=\frac{E^{2}-W(r)}{W(r)\,g_{rr}(r,0)}\,,\qquad
W(r)\equiv -g_{tt}|_{\vartheta=0}=\frac{e^{2\phi}\Delta}{\Sigma_{+}}\,,\qquad
g_{rr}|_{\vartheta=0}=\frac{e^{2\phi}\Sigma_{+}}{r^{2}}\,.
\label{eq:axial_geo}
\ee
The near-shell change of character of $W$ is displayed in Fig.~\ref{fig:effpot}.
The allowed region is $\{r:W(r)\leq E^{2}\}$ with turning points at $W(r_{*})=E^{2}$.  A probe released at rest at $r_{0}$ has $E^{2}=W(r_{0})$ and initially accelerates in the direction of decreasing $W$; the polar repulsion criterion is therefore $\partial_{r}W<0$ at $r_{0}$.  Using the identity $\mm\Sigma_{-}+\rBL\Lambda=(\rBL-\mm)\Sigma_{+}$ (and, on axis, $\Lambda=\Delta$),
\be
\partial_{r}\ln W\big|_{\vartheta=0}=2\phi'+\frac{2\sqrt{\Delta}\,\mm\Sigma_{-}}{r\Delta\Sigma_{+}}=\frac{2\sqrt{\Delta}}{r\,\Delta\,\Sigma_{+}}\,\cF(r,0)\,,
\ee
so $\partial_{r}W<0\Leftrightarrow\cF<0$ everywhere outside the outer shell: the geodesic kinematic criterion coincides identically with the static observer orthonormal criterion of the main text~(\ref{Frep}).  For supercritical $|\qq|>\mmj$, the divergence exponent analysis~(\ref{eq:rPlusDivRot}) gives $W\sim\Delta^{1-|\qq|/\mmj}\to\infty$ as $r\to r_{+}^{+}$, while $W\to 1$ at spatial infinity; $W$ therefore develops a divergent inner barrier, and a polar probe released at rest at any $r_{0}$ in the repulsive region bounces between $r_{0}$ and the matching turning point $W(r_{*})=W(r_{0})$ without reaching the singular shell.  For subcritical $|\qq|<\mmj$, $W\to 0$ at $r_{+}$ and the probe falls in (no barrier).  The polar repulsion is thus the geodesic reflection of the divergent dilaton at the FJNW-type curvature singular shell, and the threshold $|\qq|=\mmj$ is the crossing exponent that flips $W$ from regular to divergent at $r_{+}$.

\subsection{Axial null geodesic affine integral and threshold}
The axial null geodesic admits an explicit affine-parameter check.  On axis ($\vartheta=0$) with the same zero-angular-momentum condition $L=p_{\varphi}=0$, $\Lambda=\Delta$ and $\cB|_{\vartheta=0}=\Sigma_{+}/r^{2}$, so the line element collapses to a $(t,r)$ block
\be
\rd s^{2}\big|_{\vartheta=0}=-\frac{e^{2\phi}\Delta}{\Sigma_{+}}\,\rd t^{2}+\frac{e^{2\phi}\Sigma_{+}}{r^{2}}\,\rd r^{2}\,,
\label{eq:axialMetric}
\ee
with $\Sigma_{+}|_{\vartheta=0}=\rBL^{2}+j^{2}$.  A null geodesic obeys $g_{tt}\dot t^{2}+g_{rr}\dot r^{2}=0$ with conserved energy $E=-g_{tt}\dot t$, giving $\dot r^{2}=E^{2}/(-g_{tt}\,g_{rr})|_{\vartheta=0}$ and the affine parameter for an ingoing axial null ray to reach the shell from $r_{0}$
\be
\Delta\lambda=\frac{1}{E}\int_{r_{+}}^{r_{0}}\!\!\sqrt{-g_{tt}\,g_{rr}}\,\big|_{\vartheta=0}\,\rd r=\frac{1}{E}\int_{r_{+}}^{r_{0}}\!\!\frac{e^{2\phi(r)}\sqrt{\Delta(r)}}{r}\,\rd r\,.
\label{eq:nullAffine}
\ee
Near the shell, $\sqrt{\Delta}\sim 2\bar\epsilon$ ($\bar\epsilon\equiv r-r_{+}\to 0^{+}$) and $r\to r_{+}=\mmj/2$ remains finite.  On branches with axion ($0<\zeta<\pi/2$) the dominant dilaton factor diverges as $e^{2\phi}\sim\cG^{-2|\qq|/\mmj}\sin^{2}\!\zeta\sim\bar\epsilon^{-2|\qq|/\mmj}$, so
\be
\sqrt{-g_{tt}\,g_{rr}}\,\big|_{\vartheta=0}\sim\bar\epsilon^{\,1-2|\qq|/\mmj}\,.
\label{eq:nullIntegrand}
\ee
The integral~(\ref{eq:nullAffine}) is therefore convergent for $|\qq|<\mmj$ (axial photons reach the shell at finite affine parameter, terminating at the curvature singularity~(\ref{eq:rPlusDivRot})), logarithmically divergent at the threshold $|\qq|=\mmj$, and power-law divergent for $|\qq|>\mmj$ (axial photons cannot reach the shell at any finite affine parameter).  This null kinematic derivation of the threshold $|\qq|=\mmj$ is independent of the timelike effective potential $W(r)$ of~(\ref{eq:axial_geo}) and confirms that the polar repulsion threshold simultaneously shields $r_{+}$ from axial null and axial timelike probes.  On the FJNW dilaton branch ($\zeta=0$), $e^{2\phi}=\cG^{2\qq/\mmj}$ and~(\ref{eq:nullIntegrand}) becomes $\sim\bar\epsilon^{\,1+2\qq/\mmj}$; the integral diverges for $\qq<-\mmj$ only, recovering the one-sided FJNW criterion of~(\ref{eq:rPlusDivAxis}).  We make no global geodesic-completeness claim for generic nonaxial motion here.

\subsection{Derivation of the static-observer orthonormal acceleration~(\ref{Frep})}
The string frame line element~(\ref{metric}) yields $g_{tt}=-e^{2\phi}\Lambda/\Sigma_{+}$, $g_{rr}=e^{2\phi}\cB$, $g^{rr}=e^{-2\phi}\cB^{-1}$.  For the static observer $u^{\mu}=\delta^{\mu}_{t}/\sqrt{-g_{tt}}$ with $u^{r}=0$, stationarity ($\partial_{t}g_{t\sigma}=0$) gives $\Gamma^{r}_{tt}=-\tfrac{1}{2}g^{rr}\partial_{r}g_{tt}$, so
\be
a^{r}=u^{\nu}\nabla_{\nu}u^{r}=\Gamma^{r}_{tt}(u^{t})^{2}=g^{rr}\partial_{r}\ln\sqrt{-g_{tt}}\,.
\ee
The orthonormal radial component then reads
\be
a^{\hat r}=\sqrt{g_{rr}}\,a^{r}=\frac{e^{-\phi}}{\sqrt{\cB}}\!\left[\partial_{r}\phi+\tfrac{1}{2}\frac{\partial_{r}\Delta}{\Lambda}-\tfrac{1}{2}\frac{\partial_{r}\Sigma_{+}}{\Sigma_{+}}\right]\!.
\ee
From~(\ref{aux}), $\partial_{r}\rBL=\sqrt{\Delta}/r$, $\partial_{r}\Delta=2(\rBL-\mm)\sqrt{\Delta}/r$, $\partial_{r}\Sigma_{+}=2\rBL\sqrt{\Delta}/r$.  Combining the two fractions and simplifying the numerator,
\be
(\rBL-\mm)\Sigma_{+}-\rBL\Lambda=\mm\Sigma_{-}\,,
\ee
where the identity $\Lambda=\rBL^{2}-2\mm\rBL+j^{2}\cos^{2}\!\vartheta$ was used.  Factoring out $e^{-\phi}\sqrt{\Delta}/[r\sqrt{\cB}\,\Sigma_{+}\,\Lambda]$ gives
\be
a^{\hat r}=\frac{e^{-\phi}\sqrt{\Delta}}{r\sqrt{\cB}\,\Sigma_{+}\,\Lambda}\,\cF(r,\vartheta)\,,
\ee
with $\cF$ as defined in the main text~(\ref{Frep}).

\end{document}